\documentclass[aps,prb,twocolumn,groupedaddress]{revtex4-2}

\usepackage{graphicx}
\usepackage{dcolumn}
\usepackage{bm}
\usepackage{amsmath}
\usepackage{amssymb}
\usepackage{braket}
\usepackage{siunitx}
\usepackage[T1]{fontenc}
\usepackage{xcolor}

\usepackage{tikz}
\usetikzlibrary{decorations.pathmorphing,arrows.meta,calc,positioning}

\usepackage[colorlinks]{hyperref}
\usepackage[noabbrev]{cleveref}
\crefname{equation}{Eq.}{Eqs.}
\crefrangelabelformat{equation}{(#3#1#4--#5#2#6)}

\graphicspath{{.},{./fig/},{./images/}}
\DeclareGraphicsExtensions{.pdf,.png,.eps}

\begin{document}


\title{Nonlinear translational-gauge dynamics of acoustic Goldstone modes in crystals}

\author{Marco Vallone}
\email{marco.vallone@polito.it}
\affiliation{Dipartimento di Elettronica e Telecomunicazioni, Politecnico di Torino,\\
Corso Duca degli Abruzzi 24, 10129 Torino, Italy}

\date{\today}

\begin{abstract}
Acoustic phonons are the Nambu--Goldstone modes associated with broken spatial translations in crystals. We develop a nonlinear gauge-based description of their dynamics, starting from a Yang--Mills-type action constructed from the commutator of local translational derivatives. Kinematic constraints imposed by the crystalline medium restrict the gauge fields to three independent acoustic degrees of freedom, while uniform acoustic-shift symmetry and spacetime-inversion parity select the admissible nonlinear interactions. The resulting Lagrangian determines both the acoustic phonon dispersions and the cubic and quartic interaction vertices within a unified, materially constrained framework. We test these dynamics through two complementary nonlinear analyses. First, combining the resulting dispersion relations with the cubic interaction, we evaluate the kinematics and decay rate of a longitudinal acoustic phonon into two transverse acoustic phonons, finding that the channel remains kinematically open across the entire longitudinal domain with the known $k^5$ infrared scaling of the decay rate. Second, in the one-dimensional longitudinal sector, we obtain periodic Stokes-wave solutions exhibiting second-harmonic generation and no frequency shift at the leading nonlinear order.
\end{abstract}

\maketitle

\section{Introduction}
\label{sec:intro}

The Nambu--Goldstone theorem connects the spontaneous breaking of continuous symmetries to the emergence of low-energy collective excitations. In nonrelativistic condensed-matter systems, these Goldstone modes are gapless, $\omega_s(\bm{k}) \to 0$ as $k = |\bm{k}| \to 0$, where $\omega_s(\bm{k})$ denotes the angular frequency of branch $s$ at wave vector $\bm{k}$. Standard examples include spin waves in magnetically ordered Heisenberg systems~\cite{2005Kendziora_PRL} and phase modes associated with $\mathrm{U}(1)$ symmetry breaking in superconductors, whose spectrum is gapped by electromagnetic interactions via the Anderson--Higgs mechanism~\cite{2017Pracht,2017Pracht_PRB}. In crystalline solids, the periodic lattice spontaneously breaks continuous spatial translations down to a discrete subgroup, yielding three acoustic phonon branches as the corresponding Nambu--Goldstone modes in three dimensions~\cite{1976Ashcroft,1994Leutwyler_PRD,1997Leutwyler_HEPA,2002Low_PRL,2013Watanabe_PRL}.

While the Nambu--Goldstone theorem guarantees excitations of arbitrarily low energy in the thermodynamic limit, it does not determine their dispersion relations or their realization in terms of microscopic or effective fields. Moreover, when spacetime symmetries are spontaneously broken, counting independent Goldstone modes requires additional kinematic constraints~\cite{2002Low_PRL,2013Watanabe_PRL}.

At long wavelengths, acoustic Goldstone modes propagate as elastic waves whose quanta are phonons. Beyond governing lattice thermodynamics, phonons strongly influence electronic and thermal transport. In conventional superconductors, phonon exchange mediates the effective attraction responsible for Cooper pairing~\cite{1957Bardeen1_PR,1957Bardeen2_PR}, whereas electron--phonon scattering is a primary channel for carrier energy and momentum relaxation. In semiconductor heterostructures, single- and multiphonon emission similarly contributes to carrier relaxation alongside electron--electron and defect scattering~\cite{1996Kalna_PRB,2004Fan_APL,2011Sizov_JSTQE,2002Vallone_JAP,2015Vallone_PSSB}.

Symmetry-based effective field theories describe long-wavelength collective dynamics across a wide range of condensed-matter systems~\cite{2010Brauner_SYM}, including supersolids---where intertwined broken symmetries modify Goldstone-mode counting~\cite{2013Watanabe_PRL}---structural modes in perovskites and hexagonal manganites~\cite{2018Marthinsen_PRM,2020Meier_PRB}, van der Waals heterostructures~\cite{2020Koshino_PRB}, and the hydrodynamics of dissipative crystals~\cite{2022Delacretaz_PRL}.

A gauge-theoretic description of acoustic phonons was proposed by Dartora and Cabrera~\cite{2014Dartora_JPA}. By requiring the electronic Lagrangian of a crystal to be invariant under local translations, they introduced compensating gauge fields whose dynamics parallel those of gravitational gauge theories. In the linear regime, the resulting field equations describe elastic waves whose quantized modes reproduce the long-wavelength acoustic dispersion $\omega_k = c_s k$, recover standard electron--phonon couplings, and predict additional interactions between lattice strain and electronic spin currents in the presence of spin--orbit coupling.

Building on that framework, our earlier work~\cite{2019Vallone_PSSB} explored nonlinear extensions of the translational gauge formulation, incorporating gauge-field self-interactions and anharmonic phonon dynamics. That exploratory treatment, however, neither consistently isolated the three physical acoustic degrees of freedom nor eliminated spurious restoring energies associated with rigid rotations.

Here, we develop a nonlinear description of acoustic phonons in crystalline solids, starting from a Yang--Mills-type action built from the commutator of local translational derivatives. The physical properties of the crystal enter through kinematic material constraints that restrict the gauge potentials to three independent acoustic fields. The elastic response further determines which contributions to the parent action are retained: the field-strength sectors are selected to exclude quadratic restoring energies under infinitesimal rigid rotations, while uniform acoustic-shift symmetry and spacetime-inversion parity constrain the nonlinear interactions. At quadratic order, the resulting Lagrangian describes one longitudinal and two degenerate transverse gapless acoustic branches; at higher orders, it determines the cubic and quartic phonon interactions.

We test the resulting dynamics through two complementary nonlinear analyses. First, combining the predicted dispersions with the three-derivative cubic interaction, we analyze the kinematics and decay rate of a longitudinal-acoustic ($\mathrm{LA}$) phonon into two transverse-acoustic ($\mathrm{TA}$) phonons. The channel remains kinematically accessible throughout the longitudinal domain, and the infrared decay rate scales as $\Gamma_{\mathrm{LTT}} \propto \ell^2 k^5$. Second, in the one-dimensional longitudinal sector, the closed-form admissible Lagrangian yields perturbative Stokes waves with a second harmonic and no frequency shift at the leading nonlinear order.

The remainder of this paper is organized as follows. Section~\ref{sec:spacetime_gauge} develops the constrained translational-gauge formulation, derives the symmetry-allowed nonlinear phonon Lagrangian and acoustic spectrum, and examines anharmonic interactions and one-dimensional longitudinal dynamics. Section~\ref{sec:discussion} discusses the physical implications, decay kinematics, and limitations of the formulation. Section~\ref{sec:conclusions} summarizes the main results. Finally, Appendix~\ref{app:field_strength_components} presents the constrained field-strength components, while Appendix~\ref{app:decay_scaling} details the calculation of the infrared phonon decay rate.

Throughout this paper, Greek indices ($\mu,\nu,\dots = 0,1,2,3$) label local coordinates $x^\mu$ on a four-dimensional manifold $\mathcal{M}$ with metric $g_{\mu\nu}$, whereas hatted Greek indices ($\hat{\mu},\hat{\nu},\dots = \hat{0},\hat{1},\hat{2},\hat{3}$) label orthonormal tetrad frames with Minkowski metric $\eta_{\hat{\mu}\hat{\nu}} = \mathrm{diag}(+1,-1,-1,-1)$. Latin indices $i,j,k = 1,2,3$ and $\hat{i},\hat{j},\hat{k} = \hat{1},\hat{2},\hat{3}$ denote spatial coordinate and tetrad components, respectively (so that $v_{\hat{i}} = -v^{\hat{i}}$), while repeated lower spatial tetrad indices are contracted with $\delta^{\hat{i}\hat{j}}$. For acoustic applications, we set $x^0 = c_s t$, where the reference velocity scale $c_s > 0$ fixes the flat line element $\mathrm{d}s^2 = c_s^2\mathrm{d}t^2 - \mathrm{d}\bm{x}^2$; identifying this reference cone with an acoustic cone is a material input, whereas the physical longitudinal and transverse sound velocities emerge from the projected field equations.

\section{Local translational symmetry and gauge fields}
\label{sec:spacetime_gauge}

Promoting a global continuous symmetry to a local one requires introducing compensating gauge fields into the covariant derivatives so that the Lagrangian density remains invariant under spacetime-dependent transformations~\cite{1929Weyl_ZP}. In electromagnetism, for instance, invariance under local $\mathrm{U}(1)$ phase transformations introduces the four-potential $A_\mu$, whose transformation law cancels the spacetime derivatives of the local phase. For spacetime translations, however, the gauge construction involves the relation between coordinate directions and local frames rather than transformations of field components at a fixed point. In the formulation adopted here, this relation is encoded by tetrad (vierbein) fields~\cite{1984Wald_GR}.

The relevance of gravitational gauge theories to the present problem is methodological, supplying the tetrad-based geometric formalism that we subsequently adapt to acoustic dynamics in a crystalline medium. Gauge formulations of gravity, initiated by Utiyama~\cite{1956Utiyama_PR} and developed within local translational and Poincar\'e frameworks~\cite{1976Cho_PRD,1983Grensing_PRD,2018Hohmann_PRD}, model spacetime as a four-dimensional differentiable manifold $\mathcal{M}$ with local coordinates $x^\mu$. At each point $x \in \mathcal{M}$, the tangent space is spanned by a local orthonormal frame labeled by hatted Greek indices $\hat{\mu}, \hat{\nu}$ and equipped with the flat Minkowski metric $\eta_{\hat{\mu}\hat{\nu}}$.

To gauge the global spacetime translation group $\mathrm{T}(4)$, we introduce the compensating gauge connection $W_{\hat{\nu}} = R_{\hat{\nu}}{}^\mu p_\mu$, where $R_{\hat{\nu}}{}^\mu$ are the translational gauge potentials and $p_\mu = -\mathrm{i}\hbar\partial_\mu$ denote the translation generators in the coordinate representation. Minimal coupling is implemented through the gauge-covariant derivative
\begin{equation}
D_{\hat{\nu}} = \delta_{\hat{\nu}}{}^\mu\partial_\mu + W_{\hat{\nu}} = e_{\hat{\nu}}{}^\mu\partial_\mu,
\label{eq:covariant_derivative}
\end{equation}
where the inverse tetrad (frame) components $e_{\hat{\nu}}{}^\mu$ are given by
\begin{equation}
e_{\hat{\nu}}{}^\mu = \delta_{\hat{\nu}}{}^\mu - \mathrm{i}\hbar R_{\hat{\nu}}{}^\mu.
\label{eq:tetrad_field}
\end{equation}
In the flat reference configuration described by Cartesian coordinates, the local frame coincides with the coordinate basis, so that $\partial_{\hat{\nu}} = \delta_{\hat{\nu}}{}^\mu\partial_\mu$ and the gauge potentials vanish identically. Provided $e_{\hat{\nu}}{}^\mu$ is real and nonsingular across $\mathcal{M}$, it maps the coordinate basis onto the local frame, defining the local translation generators $p_{\hat{\mu}} \equiv -\mathrm{i}\hbar D_{\hat{\mu}} = e_{\hat{\mu}}{}^\nu p_\nu$. Following Cho's gauge formulation of teleparallel and translational gravity~\cite{1976Cho_PRD,2013Aldrovandi_TG}, the compensating fields $R_{\hat{\nu}}{}^\mu$ quantify the distortion of the tetrad frame relative to its flat reference state.

The dual coframe $e^{\hat{\alpha}}{}_\mu$, uniquely determined by the orthogonality and completeness relations $e^{\hat{\alpha}}{}_\mu e_{\hat{\beta}}{}^\mu = \delta^{\hat{\alpha}}{}_{\hat{\beta}}$ and $e^{\hat{\alpha}}{}_\mu e_{\hat{\alpha}}{}^\nu = \delta_\mu{}^\nu$, induces the effective spacetime metric on $\mathcal{M}$ via
\begin{equation}
g_{\mu\nu} = e^{\hat{\alpha}}{}_\mu e^{\hat{\beta}}{}_\nu \eta_{\hat{\alpha}\hat{\beta}}.
\label{eq:tetrad_metric}
\end{equation}
The translational gauge potentials thus determine both the tetrad frame and the induced metric on $\mathcal{M}$.

\subsection{Frame commutator algebra and gauge Lagrangian}
\label{sec:frame_algebra}

Although the global translation generators $p_\mu = -\mathrm{i}\hbar\partial_\mu$ commute on smooth scalar fields, their local realizations as spacetime-dependent differential operators generally possess a nonvanishing Lie bracket. For two infinitesimal local translation parameters $\epsilon^\mu(x)$ and $\zeta^\mu(x)$, the commutator of the translation operators $U_\epsilon = 1 - (\mathrm{i}/\hbar)\epsilon^\mu p_\mu$ and $U_\zeta = 1 - (\mathrm{i}/\hbar)\zeta^\mu p_\mu$ evaluates at bilinear order to
\begin{equation}
[U_\epsilon,U_\zeta] = \bigl(\epsilon^\mu\partial_\mu\zeta^\nu - \zeta^\mu\partial_\mu\epsilon^\nu\bigr)\partial_\nu.
\label{eq:commuting_rules}
\end{equation}
This nonvanishing Lie bracket originates from the spacetime dependence of the local parameters, reflecting the noncommutativity of finite coordinate-dependent shifts, whereas the global $\mathrm{T}(4)$ subgroup remains Abelian.

By the same mechanism, the translational gauge potentials $W_{\hat{\alpha}}$ defined in \cref{eq:covariant_derivative} satisfy an algebra isomorphic to \cref{eq:commuting_rules},
\begin{equation}
[W_{\hat{\alpha}}, W_{\hat{\beta}}] = -\hbar^2 \bigl( R_{\hat{\alpha}}{}^\nu \partial_\nu R_{\hat{\beta}}{}^\mu - R_{\hat{\beta}}{}^\nu \partial_\nu R_{\hat{\alpha}}{}^\mu \bigr)\partial_\mu.
\label{eq:Wcomm}
\end{equation}
The commutator of the gauge-covariant derivatives in \cref{eq:covariant_derivative} defines the translational field-strength operator $G_{\hat{\alpha}\hat{\beta}}$,
\begin{equation}
G_{\hat{\alpha}\hat{\beta}} = [D_{\hat{\alpha}},D_{\hat{\beta}}]
= \partial_{\hat{\alpha}}W_{\hat{\beta}}
-\partial_{\hat{\beta}}W_{\hat{\alpha}}
+[W_{\hat{\alpha}},W_{\hat{\beta}}],
\label{eq:G}
\end{equation}
where $\partial_{\hat{\alpha}} \equiv \delta_{\hat{\alpha}}{}^\nu \partial_\nu$ and $\partial_{\hat{\alpha}}W_{\hat{\beta}} \equiv (\partial_{\hat{\alpha}} R_{\hat{\beta}}{}^\mu)p_\mu$. Its coordinate components $G_{\hat{\alpha}\hat{\beta}} = \mathcal{G}_{\hat{\alpha}\hat{\beta}}{}^\mu\partial_\mu$ read
\begin{equation}
\begin{aligned}
\mathcal{G}_{\hat{\alpha}\hat{\beta}}{}^\mu
={}&-\mathrm{i}\hbar\bigl(\partial_{\hat{\alpha}}R_{\hat{\beta}}{}^\mu
-\partial_{\hat{\beta}}R_{\hat{\alpha}}{}^\mu\bigr)\\
&-\hbar^2\bigl(R_{\hat{\alpha}}{}^\nu\partial_\nu R_{\hat{\beta}}{}^\mu
-R_{\hat{\beta}}{}^\nu\partial_\nu R_{\hat{\alpha}}{}^\mu\bigr).
\end{aligned}
\label{eq:Gexpanded}
\end{equation}
Because the differential operators $D_{\hat{\alpha}}$ are associative, they satisfy the Jacobi identity $[D_{\hat{\alpha}},[D_{\hat{\beta}},D_{\hat{\gamma}}]] + \mathrm{cyclic\ permutations} = 0$, which yields the corresponding Bianchi identities.

To govern the dynamics of the translational gauge fields on $\mathcal{M}$, we adopt the Yang--Mills-type Lagrangian density
\begin{equation}
\mathcal{L} = -\frac{1}{4}\sqrt{|g|}\, g_{\mu\nu} \eta^{\hat{\alpha}\hat{\rho}} \eta^{\hat{\beta}\hat{\sigma}} \mathcal{G}_{\hat{\alpha}\hat{\beta}}{}^\mu \mathcal{G}_{\hat{\rho}\hat{\sigma}}{}^\nu,
\label{eq:commutator_action}
\end{equation}
where $g = \det(g_{\mu\nu})$ ensures general coordinate invariance of the action measure $\mathrm{d}^4x\sqrt{|g|}$, while $g_{\mu\nu}$ and $\eta^{\hat{\alpha}\hat{\rho}}$ contract the coordinate and local Lorentz indices, respectively. Equation~\eqref{eq:commutator_action} incorporates both the geometric nonlinearities induced by the $R_{\hat{\alpha}}{}^\mu$ dependence of $g_{\mu\nu}$ and $\sqrt{|g|}$ [\cref{eq:tetrad_field,eq:tetrad_metric}] and the algebraic self-interactions arising from $[W_{\hat{\alpha}},W_{\hat{\beta}}]$.

\subsection{Kinematic constraints and acoustic Goldstone modes}
\label{sec:restricted_dynamics}

Through the vierbein [\cref{eq:tetrad_field,eq:tetrad_metric}], the compensating field $R_{\hat{\alpha}}{}^\nu$ parameterizes the effective acoustic metric relative to the flat, undeformed crystal reference configuration. Despite the formal analogy with metric perturbations in linearized general relativity, $R_{\hat{\alpha}}{}^\nu$ acts as a translational gauge potential rather than a metric or curvature variable. Its unrestricted dynamics are governed by the Yang--Mills-type action in \cref{eq:commutator_action} rather than the Einstein--Hilbert action, whereas the acoustic sector is isolated by the material conditions introduced below.

In a crystalline solid, the spontaneous breaking of continuous spatial translations yields three gapless acoustic Nambu--Goldstone modes. We identify these three independent degrees of freedom with the gauge-field components
\begin{equation}
R_{\hat{i}}{}^0 = \frac{\mathrm{i}}{\hbar} u_{\hat{i}},
\label{eq:displacement_identification}
\end{equation}
where $u_{\hat{i}}$ denote the real, dimensionless lattice displacement fields, and $u^{\hat{0}} = 0$ excludes temporal translations. Here, the compensating potentials $R_{\hat{i}}{}^0$ serve as the fundamental dynamical variables rather than the displacement fields themselves. Following Refs.~\cite{2014Dartora_JPA,2019Vallone_PSSB}, we impose the kinematic material constraints
\begin{equation}
\begin{aligned}
R_{\hat{0}}{}^0 &= 0, \qquad R_{\hat{0}}{}^i = -R_{\hat{i}}{}^0, \\
R_{\hat{i}}{}^j &= -\ell\bigl(\partial_{\hat{i}}R_{\hat{j}}{}^0 + \partial_{\hat{j}}R_{\hat{i}}{}^0\bigr),
\end{aligned}
\label{eq:material_R_constraints}
\end{equation}
where $\ell$ is a characteristic real material length scale. The last relation enforces the symmetry $R_{\hat{i}}{}^j = R_{\hat{j}}{}^i$ and corresponds to the linearized elastic strain tensor $\varepsilon_{ij} = \frac{1}{2}(\partial_i u_j + \partial_j u_i)$ up to the scale factor $\ell$. From \cref{eq:tetrad_field,eq:tetrad_metric,eq:displacement_identification}, the spatial metric to leading order in the displacements reads $g_{ij} = -\delta_{ij} - 2\ell(\partial_i u_j + \partial_j u_i) + \mathcal{O}(u^2)$. Under an infinitesimal rigid spatial rotation, $u_{\hat{i}} = \Omega_{ij}x^j$ with $\Omega_{ij} = -\Omega_{ji}$, the antisymmetry of $\Omega_{ij}$ yields $R_{\hat{i}}{}^j = 0$, leaving $g_{ij}$ invariant to linear order.

Substituting \cref{eq:material_R_constraints} into the field-strength components in \cref{eq:Gexpanded} organizes $\mathcal{G}_{\hat{\alpha}\hat{\beta}}{}^\mu$ into four spacetime index families ($\mathcal{G}_{\hat{0}\hat{i}}{}^0$, $\mathcal{G}_{\hat{0}\hat{i}}{}^j$, $\mathcal{G}_{\hat{i}\hat{j}}{}^0$, and $\mathcal{G}_{\hat{i}\hat{j}}{}^k$), listed explicitly in Appendix~\ref{app:field_strength_components}. These constrained components retain the full commutator contribution, while the tetrad and induced metric follow from the same kinematic constraints.

\subsection{Rotational objectivity and effective acoustic Lagrangian}
\label{sec:effective_lagrangian}

To express the field-strength contractions in the local frame, we convert the upper coordinate index according to $\mathcal{G}_{\hat{\alpha}\hat{\beta}}{}^{\hat{\gamma}} = e^{\hat{\gamma}}{}_{\mu}\mathcal{G}_{\hat{\alpha}\hat{\beta}}{}^\mu$. The gauge Lagrangian of \cref{eq:commutator_action} then decomposes into temporal, symmetric and antisymmetric spatial, and purely spatial sectors:
\begin{equation}
\begin{aligned}
\mathcal{L}_g ={}& \frac{\sqrt{|g|}}{2}\Bigg[
\sum_{\hat{i}}\bigl(\mathcal{G}_{\hat{0}\hat{i}}{}^{\hat{0}}\bigr)^2
-\frac{1}{4}\sum_{\hat{i},\hat{j}}
\bigl(\mathcal{G}_{\hat{0}\hat{i}}{}^{\hat{j}}
 +\mathcal{G}_{\hat{0}\hat{j}}{}^{\hat{i}}\bigr)^2 \\
&-\frac{1}{4}\sum_{\hat{i},\hat{j}}
\bigl(\mathcal{G}_{\hat{0}\hat{i}}{}^{\hat{j}}
 -\mathcal{G}_{\hat{0}\hat{j}}{}^{\hat{i}}\bigr)^2
-\sum_{\hat{i}<\hat{j}}g_{\mu\nu}
\mathcal{G}_{\hat{i}\hat{j}}{}^\mu
\mathcal{G}_{\hat{i}\hat{j}}{}^\nu\Bigg].
\end{aligned}
\label{eq:material_gauge_decomposition}
\end{equation}
Under an infinitesimal static rigid rotation $R_{\hat{i}}{}^0 = (\mathrm{i}/\hbar)\Omega_{ij}x^j$, both the temporal components and the symmetric spatial combination in \cref{eq:material_gauge_decomposition} vanish to linear order, whereas the antisymmetric spatial term assigns a nonzero restoring energy to the rotation. The purely spatial field-strength sector also contributes to the rotational restoring energy through $\mathcal{G}_{\hat{i}\hat{j}}{}^0$. We therefore retain only the temporal and symmetric spatial sectors as a minimal rotationally admissible projection. To separate the linear derivative terms ($\mathrm{lin}$) from the nonlinear self-coupling contributions ($\mathrm{s}$) arising from $[W_{\hat{\alpha}},W_{\hat{\beta}}]$ in \cref{eq:Gexpanded} prior to expanding the tetrad or metric, we decompose the local-frame field strength into two parts,
\begin{equation}
\begin{aligned}
\mathcal{G}_{\hat{0}\hat{i}}{}^{\mathrm{lin},\hat{\alpha}}
={}&-\mathrm{i}\hbar\,e^{\hat{\alpha}}{}_{\mu}
\bigl(\partial_{\hat{0}}R_{\hat{i}}{}^\mu
 -\partial_{\hat{i}}R_{\hat{0}}{}^\mu\bigr),\\
\mathcal{G}_{\hat{0}\hat{i}}{}^{\mathrm{s},\hat{\alpha}}
={}&-\hbar^2 e^{\hat{\alpha}}{}_{\mu}
\bigl(R_{\hat{0}}{}^\nu\partial_\nu R_{\hat{i}}{}^\mu
-R_{\hat{i}}{}^\nu\partial_\nu R_{\hat{0}}{}^\mu\bigr),
\end{aligned}
\label{eq:material_gauge_two_parts}
\end{equation}
so that the projected phonon Lagrangian density reads
\begin{equation}
\begin{aligned}
\mathcal{L}_{\mathrm{ph}}
={}&\frac{\sqrt{|g|}}{2}\Bigg\{
\sum_{\hat{i}}
\bigl(\mathcal{G}_{\hat{0}\hat{i}}{}^{\mathrm{lin},\hat{0}}+\mathcal{G}_{\hat{0}\hat{i}}{}^{\mathrm{s},\hat{0}}\bigr)^2\\
&-\frac{1}{4}\sum_{\hat{i},\hat{j}}
\bigl(\mathcal{G}_{\hat{0}\hat{i}}{}^{\mathrm{lin},\hat{j}}
+\mathcal{G}_{\hat{0}\hat{j}}{}^{\mathrm{lin},\hat{i}} +
\mathcal{G}_{\hat{0}\hat{i}}{}^{\mathrm{s},\hat{j}}
+\mathcal{G}_{\hat{0}\hat{j}}{}^{\mathrm{s},\hat{i}}\bigr)^2
\Bigg\}.
\end{aligned}
\label{eq:material_projected_action}
\end{equation}
In coordinate components, $\mathcal{G}^{\mathrm{lin}}$ and $\mathcal{G}^{\mathrm{s}}$ are linear and quadratic in the gauge potentials, respectively, and each involves derivatives of $R_{\hat{\alpha}}{}^\mu$. Their products contribute at quadratic, cubic, and quartic field order, while the field dependence of the coframe $e^{\hat{\alpha}}{}_\mu$ and $\sqrt{|g|}$ supplies additional geometric nonlinearities.

The equations of motion of the acoustic sector follow by varying the material action---subject to the symmetry-admissibility prescription specified below---with respect to the three independent fields $R_{\hat{i}}{}^0$. Because the resulting Lagrangian depends on field derivatives up to second order, including mixed spacetime derivatives, the generalized Euler--Lagrange equations apply.

\subsection{Linearized acoustic spectrum in three dimensions}
\label{sec:linear_spectrum}

At quadratic field order, neither the commutator self-coupling nor the field dependence of the coframe and volume element contributes. Evaluating the projected Lagrangian $\mathcal{L}_{\mathrm{ph}}$ [\cref{eq:material_projected_action}] on the flat reference background ($e^{\hat{\alpha}}{}_{\mu} = \delta^{\hat{\alpha}}{}_{\mu}$, $\sqrt{|g|} = 1$) therefore yields the free quadratic phonon Lagrangian $\mathcal{L}_{\mathrm{ph}}^{(2)}$, whose bulk field equations are unaffected by the subsequent admissibility prescription. Specifically, $\mathcal{L}_{\mathrm{ph}}^{(2)}$ comprises two-derivative kinetic and elastic terms together with four-derivative terms proportional to $\ell^2$, whereas the three-derivative cross term reduces to a total time derivative and drops out of the bulk equations of motion. Requiring the quadratic action $S_{\mathrm{ph}}^{(2)} = \int \mathrm{d}^4x\,\mathcal{L}_{\mathrm{ph}}^{(2)}$ to be stationary with respect to $R_{\hat{i}}{}^0$ gives the linearized field equations
\begin{equation}
\begin{aligned}
\partial_{\hat{0}}^2 R_{\hat{i}}{}^0
&-\frac{1}{2}\partial_{\hat{j}}\bigl(
 \partial_{\hat{j}}R_{\hat{i}}{}^0 + \partial_{\hat{i}}R_{\hat{j}}{}^0\bigr) \\
&+ 2\ell^2\partial_{\hat{0}}^2\partial_{\hat{j}}\bigl(
 \partial_{\hat{j}}R_{\hat{i}}{}^0 + \partial_{\hat{i}}R_{\hat{j}}{}^0\bigr) = 0,
\end{aligned}
\label{eq:material_linear_EL}
\end{equation}
where the fourth-order mixed derivative introduces a wave-number dependence of the acoustic phase velocities controlled by the material parameter $\ell$. Substituting the plane-wave ansatz $R_{\hat{i}}{}^0 \propto \exp[\mathrm{i}(\bm{k}\cdot\bm{x}-\omega t)]$ yields one longitudinal-acoustic ($\mathrm{LA}$) branch and two degenerate transverse-acoustic ($\mathrm{TA}$) branches with dispersion relations
\begin{equation}
\omega_{\mathrm{LA}}^2 = \frac{c_s^2 k^2}{1-4\ell^2 k^2}, \qquad
\omega_{\mathrm{TA}}^2 = \frac{c_s^2 k^2}{2(1-2\ell^2 k^2)}.
\label{eq:material_linear_dispersions}
\end{equation}
Both branches are gapless, as required for acoustic Goldstone excitations, and possess real frequencies across $|\ell|k < 1/2$. In the infrared limit $|\ell| k \ll 1$, the dispersions are linear, and the long-wavelength phase velocities $v_L = c_s$ and $v_T = c_s/\sqrt{2}$ satisfy $v_L/v_T = \sqrt{2}$, a ratio fixed by the symmetric field-strength contraction in \cref{eq:material_projected_action}.


\subsection{Symmetry-admissible anharmonic interactions}
\label{sec:cubic_interactions}

For the acoustic sector considered here, we impose simultaneous spacetime-inversion parity ($x^\mu \to -x^\mu$, $u_{\hat{i}} \to -u_{\hat{i}}$), retaining operators with $n+D$ even, where $n$ and $D$ denote the field and derivative orders, respectively. We also impose uniform acoustic-shift invariance ($u_{\hat{i}} \to u_{\hat{i}} + a_{\hat{i}}$) by requiring each field in a retained local operator to enter solely through its derivatives ($D \ge n$). This termwise derivative prescription is motivated by the Adler soft-phonon condition and enforces derivative couplings at the level of the retained local vertices. It guarantees uniform-shift invariance, while being more restrictive than requiring invariance of the action modulo total derivatives~\cite{1994Leutwyler_PRD,1997Leutwyler_HEPA,2010Brauner_SYM}.
Applying this admissibility prescription to \cref{eq:material_projected_action} eliminates undifferentiated occurrences of $R_{\hat{i}}{}^0$ (and $R_{\hat{0}}{}^i = -R_{\hat{i}}{}^0$) from both the commutator contributions and the geometric factors $e^{\hat{\alpha}}{}_\mu$ and $\sqrt{|g|}$. In the self-coupling field strength $\mathcal{G}_{\hat{0}\hat{i}}{}^{\mathrm{s},\mu}$, the sole surviving contribution is the spatial-gradient contraction proportional to $R_{\hat{i}}{}^l \propto \ell\,\partial u$, which defines the admissible self-coupling component
\begin{equation}
\begin{aligned}
\mathcal{G}_{\hat{0}\hat{i}}{}^{\mathrm{s,adm},j}
={}& -\hbar^2 R_{\hat{i}}{}^l \partial_{\hat{l}} R_{\hat{j}}{}^0 \\
={}& \hbar^2 \ell \bigl(\partial_{\hat{i}}R_{\hat{l}}{}^0 + \partial_{\hat{l}}R_{\hat{i}}{}^0\bigr)\partial_{\hat{l}}R_{\hat{j}}{}^0.
\end{aligned}
\label{eq:G_self_admissible}
\end{equation}
The second equality follows from the material constraints [\cref{eq:material_R_constraints}], whereas $\mathcal{G}_{\hat{0}\hat{i}}{}^{\mathrm{s,adm},0} = 0$. Factoring out the coframe components via $\mathcal{G}_{\hat{0}\hat{i}}{}^{\mathrm{lin},\hat{\alpha}} = e^{\hat{\alpha}}{}_{\mu}\mathcal{G}_{\hat{0}\hat{i}}{}^{\mathrm{lin},\mu}$ and $\mathcal{G}_{\hat{0}\hat{i}}{}^{\mathrm{s,adm},\hat{\alpha}} = e^{\hat{\alpha}}{}_{j}\mathcal{G}_{\hat{0}\hat{i}}{}^{\mathrm{s,adm},j}$, the symmetry-projected phonon Lagrangian organizes according to powers of the linear and self-coupling field strengths (prior to expanding the admissible geometric factors) as
\begin{equation}
\mathcal{L}_{\mathrm{ph}}^{\mathrm{adm}}
=\mathcal{L}_{\mathrm{ph}}^{\lbrack 2\rbrack}
+\mathcal{L}_{\mathrm{ph}}^{\lbrack 3\rbrack}
+\mathcal{L}_{\mathrm{ph}}^{\lbrack 4\rbrack}.
\label{eq:material_comm_order_sum}
\end{equation}
Here, the bracketed superscripts $[2]$, $[3]$, and $[4]$ label contributions involving, respectively, two linear field-strength components, one linear and one self-coupling component, and two self-coupling components; they do not denote the total field order, as the coframe and volume-element factors remain unexpanded. Specifically, $\mathcal{L}_{\mathrm{ph}}^{[2]}$ retains the structure of \cref{eq:material_projected_action} with $\mathcal{G}^{\mathrm{s}} = 0$, whereas the mixed and quadratic self-coupling contributions read
\begin{equation}
\begin{aligned}
\mathcal{L}_{\mathrm{ph}}^{\lbrack 3\rbrack}
={}&-\frac{\sqrt{|g|}}{4}\sum_{\hat{i},\hat{j}}
\Bigl(e^{\hat{j}}{}_{\mu}\mathcal{G}_{\hat{0}\hat{i}}{}^{\mathrm{lin},\mu}
+e^{\hat{i}}{}_{\mu}\mathcal{G}_{\hat{0}\hat{j}}{}^{\mathrm{lin},\mu}\Bigr)\\
&\qquad\times\Bigl(e^{\hat{j}}{}_{k}\mathcal{G}_{\hat{0}\hat{i}}{}^{\mathrm{s,adm},k}
+e^{\hat{i}}{}_{k}\mathcal{G}_{\hat{0}\hat{j}}{}^{\mathrm{s,adm},k}\Bigr),
\end{aligned}
\label{eq:material_cubic_gradient}
\end{equation}
and
\begin{equation}
\begin{aligned}
\mathcal{L}_{\mathrm{ph}}^{\lbrack 4\rbrack}
={}&-\frac{\sqrt{|g|}}{8}\sum_{\hat{i},\hat{j}}
\Bigl(e^{\hat{j}}{}_{k}\mathcal{G}_{\hat{0}\hat{i}}{}^{\mathrm{s,adm},k}
+e^{\hat{i}}{}_{k}\mathcal{G}_{\hat{0}\hat{j}}{}^{\mathrm{s,adm},k}\Bigr)\\
&\qquad\times\Bigl(e^{\hat{j}}{}_{l}\mathcal{G}_{\hat{0}\hat{i}}{}^{\mathrm{s,adm},l}
+e^{\hat{i}}{}_{l}\mathcal{G}_{\hat{0}\hat{j}}{}^{\mathrm{s,adm},l}\Bigr).
\end{aligned}
\label{eq:material_comm_quartic}
\end{equation}
The retained self-coupling in \cref{eq:G_self_admissible} is proportional to $\ell$ and quadratic in the first derivatives of $R_{\hat{i}}{}^0$. Its product with the linear field strength enters the cubic sector, whereas its square enters the quartic sector; the complete vertices at fixed field order also incorporate the coframe and volume-element contributions.

We next isolate the leading cubic interaction governing acoustic-phonon decay. At three-derivative order, the contribution from \cref{eq:material_cubic_gradient} cancels the corresponding coframe correction from $\mathcal{L}_{\mathrm{ph}}^{[2]}$ upon symmetric spatial contraction. The surviving term originates from the linear expansion of $\sqrt{|g|}$ and reads
\begin{equation}
\begin{aligned}
\mathcal{L}_{\mathrm{ph},D=3}^{(3)}
={}&\mathrm{i}\hbar^3\ell
\biggl(\sum_{\hat{l}}\partial_{\hat{l}}R_{\hat{l}}{}^0\biggr)
\Biggl\lbrack\sum_{\hat{i}}(\partial_{\hat{0}}R_{\hat{i}}{}^0)^2\\
&-\frac{1}{4}\sum_{\hat{i},\hat{j}}
\bigl(\partial_{\hat{i}}R_{\hat{j}}{}^0
+\partial_{\hat{j}}R_{\hat{i}}{}^0\bigr)^2\Biggr\rbrack.
\end{aligned}
\label{eq:material_complete_cubic}
\end{equation}
Equation~\eqref{eq:material_complete_cubic} represents the complete three-derivative cubic vertex, including all geometric contributions, where the superscript $(3)$ denotes the field order while $[3]$ labels the mixed linear/self-coupling sector prior to expanding the geometric factors in \cref{eq:material_comm_order_sum}.

As a direct physical application of the cubic interaction in \cref{eq:material_complete_cubic}, we evaluate the zero-temperature decay rate of a longitudinal-acoustic ($\mathrm{LA}$) phonon into two transverse-acoustic ($\mathrm{TA}$) phonons, $\mathrm{LA}(\bm{k}) \to \mathrm{TA}(\bm{q}_+) + \mathrm{TA}(\bm{q}_-)$, with $\bm{q}_+ + \bm{q}_- = \bm{k}$. We first verify that this decay channel is kinematically open in the long-wavelength limit. Consider an incoming $\mathrm{LA}$ phonon with wave vector $\bm{k} = k\hat{\bm{x}}$ and polarization $\hat{\bm{e}}_{\mathrm{L}} = \hat{\bm{x}}$, decaying into two outgoing $\mathrm{TA}$ phonons with wave vectors $\bm{q}_\pm = \tfrac{k}{2}(\hat{\bm{x}} \pm \hat{\bm{y}})$, both polarized along $\hat{\bm{z}}$. These wave vectors satisfy $\bm{q}_+ + \bm{q}_- = \bm{k}$ with $q_+ = q_- = k/\sqrt{2}$. From the acoustic dispersion relations derived in Sec.~\ref{sec:linear_spectrum}, whose infrared limits are $\omega_{\mathrm{LA}}(k) = c_s k[1 + \mathcal{O}(\ell^2k^2)]$ and $\omega_{\mathrm{TA}}(q) = (c_s q/\sqrt{2})[1 + \mathcal{O}(\ell^2q^2)]$, we obtain
\begin{equation}
\begin{aligned}
\omega_{\mathrm{TA}}(q_+) + \omega_{\mathrm{TA}}(q_-) ={}& c_s k[1 + \mathcal{O}(\ell^2k^2)] \\
={}& \omega_{\mathrm{LA}}(k) + \mathcal{O}(c_s\ell^2k^3),
\end{aligned}
\nonumber
\end{equation}
demonstrating that both momentum and energy are conserved at leading order in the infrared regime. Moreover, the cubic vertex couples these modes dynamically: in \cref{eq:material_complete_cubic}, the incoming longitudinal mode enters through $\partial_{\hat{l}}R_{\hat{l}}{}^0$, whereas the outgoing transverse modes enter through the quadratic factor. For the chosen polarizations, the spatial contraction vanishes because $\bm{q}_+ \cdot \bm{q}_- = 0$, while the temporal contraction remains nonzero, yielding a nonvanishing decay amplitude.

For a general final-state configuration, let $\bm{q}_+ = \bm{q}$ and $\bm{q}_- = \bm{k} - \bm{q}$, with $\sigma_\pm$ labeling the transverse polarizations. Between the initial state $\ket{i} = \ket{\mathrm{LA}(\bm{k})}$ and the final state $\ket{f_{\bm{q},\sigma_+,\sigma_-}} = \ket{\mathrm{TA}(\bm{q}_+,\sigma_+),\mathrm{TA}(\bm{q}_-,\sigma_-)}$, the transition matrix element reads
\begin{equation}
\mathcal{M}_{\sigma_+\sigma_-}(\bm{k},\bm{q})
= \bra{f_{\bm{q},\sigma_+,\sigma_-}}
\hat{H}_{\mathrm{ph}}^{(3)}\ket{i},
\label{eq:decay_LTT_matrix_element}
\end{equation}
where the cubic interaction Hamiltonian $\hat{H}_{\mathrm{ph}}^{(3)}$ is obtained from \cref{eq:material_complete_cubic} via Legendre transformation. According to Fermi's golden rule~\cite{1961Pokrovskii_Dykhne_JETP}, the decay rate is given by
\begin{equation}
\begin{aligned}
\Gamma_{\mathrm{LTT}}(k)
={}&\frac{2\pi}{\hbar\,2!}
\sum_{\bm{q},\sigma_+,\sigma_-}
\bigl|\mathcal{M}_{\sigma_+\sigma_-}
(\bm{k},\bm{q})\bigr|^2\\
&\times\delta\!\Bigl[
\hbar\omega_{\mathrm{LA}}(k)
-\hbar\omega_{\mathrm{TA}}(q_+)
-\hbar\omega_{\mathrm{TA}}(q_-)
\Bigr],
\end{aligned}
\label{eq:decay_rate_golden_rule}
\end{equation}
where the symmetry factor $1/2!$ accounts for the two indistinguishable outgoing bosons.

At long wavelengths, the three derivatives in \cref{eq:material_complete_cubic} produce a vertex scaling as $\ell k^3$. Within a quantization volume $V$, each canonically normalized phonon operator carries a factor proportional to $(V\omega)^{-1/2}$, contributing an overall prefactor $[V^3\omega_{\mathrm{LA}}(k)\omega_{\mathrm{TA}}(q_+)\omega_{\mathrm{TA}}(q_-)]^{-1/2}$ to the amplitude. Since all three frequencies scale linearly with $k$ on shell, the squared matrix element scales as
\begin{equation}
\bigl|\mathcal{M}_{\sigma_+\sigma_-}(\bm{k},\bm{q})\bigr|^2
\sim \frac{\ell^2 k^3}{V},
\label{eq:decay_rate_matrix_scaling}
\end{equation}
up to $k$-independent material and polarization factors. In three dimensions, taking the continuum limit $\sum_{\bm{q}} \to V\int \mathrm{d}^3q/(2\pi)^3$ and rescaling $\bm{q} = k\bm{p}$, the integration measure contributes a factor of $k^3$ while the energy-conserving delta function contributes $k^{-1}$. The available phase space therefore scales as $Vk^2$, yielding the infrared decay rate
\begin{equation}
\Gamma_{\mathrm{LTT}}(k)
\propto \ell^2 k^3\,k^2
= \ell^2 k^5.
\label{eq:decay_rate_k5}
\end{equation}
The full canonical derivation and polarization sum are detailed in Appendix~\ref{app:decay_scaling}.

\subsection{Nonlinear longitudinal dynamics}
\label{sec:1d_exact}

For a longitudinal mode propagating along $x^1$, the sole independent gauge component is $R_{\hat{1}}{}^0(x^0,x^1)$. The kinematic constraints of \cref{eq:material_R_constraints} yield $R_{\hat{0}}{}^1 = -R_{\hat{1}}{}^0$ and $R_{\hat{1}}{}^1 = -2\ell R_{\hat{1},\hat{1}}{}^0$, with all transverse components vanishing. Here and below, a comma denotes partial differentiation in the Cartesian reference coordinates, $\partial_{\hat{\alpha}} = \delta_{\hat{\alpha}}{}^\mu\partial_\mu$; for example, $R_{\hat{1},\hat{0}\hat{1}}{}^0 = \partial_{\hat{0}}\partial_{\hat{1}} R_{\hat{1}}{}^0$.

Applying the symmetry-admissibility projection of Sec.~\ref{sec:cubic_interactions} directly to the longitudinal sector eliminates all nonderivative occurrences of $R_{\hat{1}}{}^0$ from both the tetrad and the commutator self-interaction. In the admissible action, the retained longitudinal inverse-tetrad factors are represented by the strain-dependent diagonal form
\begin{equation}
\lbrack e_{\hat{\alpha}}{}^\mu\rbrack_{0,1}^{\mathrm{adm}}
=\begin{pmatrix}
1 & 0\\
0 & 1+2\mathrm{i}\hbar\ell R_{\hat{1},\hat{1}}{}^0
\end{pmatrix},
\label{eq:material_1d_tetrad}
\end{equation}
whose determinant $d_{\mathrm{adm}} = 1 + 2\mathrm{i}\hbar\ell R_{\hat{1},\hat{1}}{}^0$ fixes the volume element $\sqrt{|g|} = d_{\mathrm{adm}}^{-1}$ on the branch $d_{\mathrm{adm}} > 0$ (whereas the unprojected tetrad also includes off-diagonal $\pm\mathrm{i}\hbar R_{\hat{1}}{}^0$ entries). Similarly, the admissible self-coupling field strength of \cref{eq:G_self_admissible} simplifies to the single spatial component $\mathcal{G}_{\hat{0}\hat{1}}{}^{\mathrm{s,adm},1} = 2\hbar^2\ell(R_{\hat{1},\hat{1}}{}^0)^2$, whereas $\mathcal{G}_{\hat{0}\hat{1}}{}^{\mathrm{s,adm},0} = 0$.

Using the inverse of $\lbrack e_{\hat{\alpha}}{}^\mu\rbrack_{0,1}^{\mathrm{adm}}$ for the local-frame contractions yields the closed-form, symmetry-admissible one-dimensional Lagrangian
\begin{equation}
\begin{aligned}
\mathcal{L}_{\mathrm{1D}}^{\mathrm{adm}}
={}&\frac{\hbar^2}{2}\Biggl\{
-\frac{\bigl(R_{\hat{1},\hat{0}}{}^0\bigr)^2}{d_{\mathrm{adm}}} \\
&+\frac{\bigl\lbrack R_{\hat{1},\hat{1}}{}^0
-2\ell R_{\hat{1},\hat{0}\hat{1}}{}^0
+2\mathrm{i}\hbar\ell\bigl(R_{\hat{1},\hat{1}}{}^0\bigr)^2\bigr\rbrack^2}
{d_{\mathrm{adm}}^3}
\Biggr\},
\end{aligned}
\label{eq:material_1d_exact_lagrangian}
\end{equation}
which depends exclusively on spacetime derivatives of $R_{\hat{1}}{}^0$ and is therefore manifestly invariant under uniform displacement shifts ($R_{\hat{1}}{}^0 \to R_{\hat{1}}{}^0 + \mathrm{const}$) at all nonlinear orders. Expanding \cref{eq:material_1d_exact_lagrangian} to quadratic order in $R_{\hat{1}}{}^0$ gives the harmonic longitudinal Lagrangian
\begin{equation}
\mathcal{L}_{\mathrm{1D}}^{(2)} = \frac{\hbar^2}{2}\Bigl\lbrack-(R_{\hat{1},\hat{0}}{}^0)^2+(R_{\hat{1},\hat{1}}{}^0 - 2\ell R_{\hat{1},\hat{0}\hat{1}}{}^0)^2
\Bigr\rbrack.
\label{eq:material_1d_quadratic}
\end{equation}
The mixed term in \cref{eq:material_1d_quadratic} is a total time derivative; varying the action yields the linearized equation of motion
\begin{equation}
\partial_{\hat{0}}^2 R_{\hat{1}}{}^0 - \partial_{\hat{1}}^2 R_{\hat{1}}{}^0
+ 4\ell^2 \partial_{\hat{0}}^2 \partial_{\hat{1}}^2 R_{\hat{1}}{}^0 = 0,
\label{eq:material_1d_linear_EL}
\end{equation}
in agreement with the longitudinal sector of \cref{eq:material_linear_EL} and the $\mathrm{LA}$ dispersion in \cref{eq:material_linear_dispersions}.

To isolate the anharmonic couplings governing longitudinal-wave dynamics, we expand the exact shift-invariant Lagrangian in \cref{eq:material_1d_exact_lagrangian} to cubic order in the independent field $R_{\hat{1}}{}^0$ at fixed $\ell$. Because the four-derivative cubic terms reduce to total derivatives and drop out of the field equation, the leading cubic interaction enters at three-derivative order:
\begin{equation}
\begin{aligned}
\mathcal{L}_{\mathrm{1D},D=3}^{\mathrm{adm},(3)}
={}&\mathrm{i}\hbar^3\ell R_{\hat{1},\hat{1}}{}^0
\Bigl[
\bigl(R_{\hat{1},\hat{0}}{}^0\bigr)^2
-\bigl(R_{\hat{1},\hat{1}}{}^0\bigr)^2
\Bigr]\\
={}&\ell(\partial_1 u_1)
\Bigl[
(\partial_0 u_1)^2-(\partial_1 u_1)^2
\Bigr].
\end{aligned}
\label{eq:material_1d_cubic}
\end{equation}
At the same cubic order, the expansion also generates a five-derivative contribution,
\begin{equation}
\mathcal{L}_{\mathrm{1D},D=5}^{\mathrm{adm},(3)}
=-12\mathrm{i}\hbar^3\ell^3
R_{\hat{1},\hat{1}}{}^0
\bigl(R_{\hat{1},\hat{0}\hat{1}}{}^0\bigr)^2,
\label{eq:material_1d_cubic_D5}
\end{equation}
which is subleading in the long-wavelength regime. Both vertices are automatically resummed within the closed-form Lagrangian of \cref{eq:material_1d_exact_lagrangian}, which we employ below to analyze nonlinear periodic waves.

To characterize finite-amplitude propagation in this sector, we consider periodic traveling-wave solutions of the exact nonlinear longitudinal field equation. Because $\mathcal{L}_{\mathrm{1D}}^{\mathrm{adm}}$ depends on $R_{\hat{1}}{}^0$ solely through its derivatives, the resulting Euler--Lagrange equation assumes conservation form; consequently, its spatial average over one period vanishes identically, precluding any mean-value solvability obstruction. Small-amplitude periodic solutions follow from the Stokes expansion~\cite{1974Whitham}
\begin{equation}
\begin{aligned}
R_{\hat{1}}{}^0
&= -\frac{\mathrm{i}}{\hbar}
\bigl[ A\cos\xi + A^2 f^{(2)}(\xi)
+ \mathcal{O}(A^3)\bigr], \\
\xi &= kx^1 - \omega t,
\end{aligned}
\label{eq:material_1d_stokes_ansatz}
\end{equation}
where $A$ is a real dimensionless amplitude, $f^{(2)}(\xi)$ denotes a real $2\pi$-periodic correction, and $\omega \to \omega_{\mathrm{LA}}(k)$ in the linear limit $A \to 0$. With $x^0 = c_s t$, the reference derivatives act as $\partial_{\hat{0}} = -(\omega/c_s)\,\mathrm{d}/\mathrm{d}\xi$ and $\partial_{\hat{1}} = k\,\mathrm{d}/\mathrm{d}\xi$. Inserting this expansion into the exact Euler--Lagrange equation and matching the nonconstant Fourier harmonics through cubic order in $A$ yields
\begin{equation}
\begin{aligned}
f^{(2)}(\xi)&=-\frac{\ell k}{2}\sin(2\xi),\\
\frac{\Delta\omega}{\omega_{\mathrm{LA}}(k)}&=\mathcal{O}(A^4),
\end{aligned}
\label{eq:material_1d_stokes_shift}
\end{equation}
where $\Delta\omega\equiv\omega-\omega_{\mathrm{LA}}(k)$, valid for $|\ell|k<1/2$ and sufficiently small $A$. The nonlinear interaction thus generates a second harmonic at $\mathcal{O}(A^2)$ while leaving the wave frequency unshifted through this order. By contrast, the nonderivative couplings retained in the unprojected constrained action can induce a nonzero period average of the Euler--Lagrange equation, precluding periodic solutions.

\section{Discussion}
\label{sec:discussion}

The results obtained above provide two complementary tests of the symmetry-admissible acoustic action: the $\mathrm{LA}\to\mathrm{TA}+\mathrm{TA}$ decay process, which probes both the harmonic dispersions and the cubic interaction, and nonlinear longitudinal-wave propagation, which tests the closed-form one-dimensional action. We examine these physical consequences before addressing the structural role of the material projection and its symmetries.

\begin{figure}[t]
\includegraphics[width=\columnwidth]{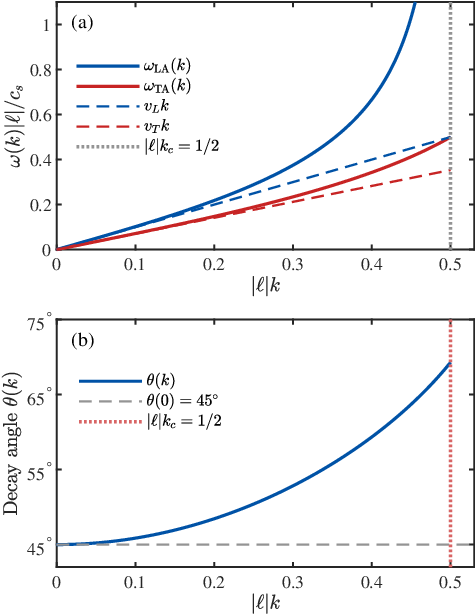}
\caption{(a) Dimensionless acoustic dispersion branches $\omega_{\mathrm{LA}}(k)|\ell|/c_s$ and $\omega_{\mathrm{TA}}(k)|\ell|/c_s$ [\cref{eq:material_linear_dispersions}] up to the cutoff $|\ell|k_c = 1/2$, shown alongside their linear long-wavelength asymptotes (dashed lines). (b) Opening angle $\theta(k)$ [\cref{eq:decay_opening_angle}] for symmetric $\mathrm{LA}(\bm{k}) \to \mathrm{TA}(\bm{q}_+) + \mathrm{TA}(\bm{q}_-)$ decay, increasing from $\theta(0) = 45^\circ$ to $\theta(k_c^-) \approx 69.3^\circ$.}
\label{fig:dispersion_kinematics}
\end{figure}

We first assess whether the predicted spectrum permits the decay of a longitudinal phonon into two transverse phonons. Figure~\ref{fig:dispersion_kinematics}(a) shows the acoustic branches as functions of $|\ell|k$. The LA dispersion steepens with increasing wave vector and exhibits a pole at $|\ell|k_c = 1/2$, whereas the TA branch remains real for $|\ell|k < 1/\sqrt{2}$. This longitudinal pole signals the breakdown of the effective continuum description near $k_c$ rather than a physical divergence of the lattice frequency.

For symmetric decay $\mathrm{LA}(\bm{k}) \to \mathrm{TA}(\bm{q}_+) + \mathrm{TA}(\bm{q}_-)$, let $\theta$ denote the angle between either outgoing wave vector and $\bm{k}$. Momentum conservation requires $|\bm{q}_+| = |\bm{q}_-| = k/(2\cos\theta)$. Combining this relation with exact energy conservation, $\omega_{\mathrm{LA}}(k) = 2\omega_{\mathrm{TA}}(q)$, yields
\begin{equation}
\cos^2\theta(k) = \frac{1 - 3\ell^2 k^2}{2},
\label{eq:decay_opening_angle}
\end{equation}
plotted in Fig.~\ref{fig:dispersion_kinematics}(b). The infrared limit $\theta(0) = 45^\circ$ reproduces the orthogonal outgoing-wave geometry of Sec.~\ref{sec:cubic_interactions}. The angle increases monotonically toward $\theta(k_c^-) = \arccos(1/\sqrt{8}) \approx 69.3^\circ$. Thus, the predicted LA and TA dispersions keep this decay channel kinematically open across the entire longitudinal domain $|\ell|k < 1/2$.

\begin{figure}[t]
\includegraphics[width=\columnwidth]{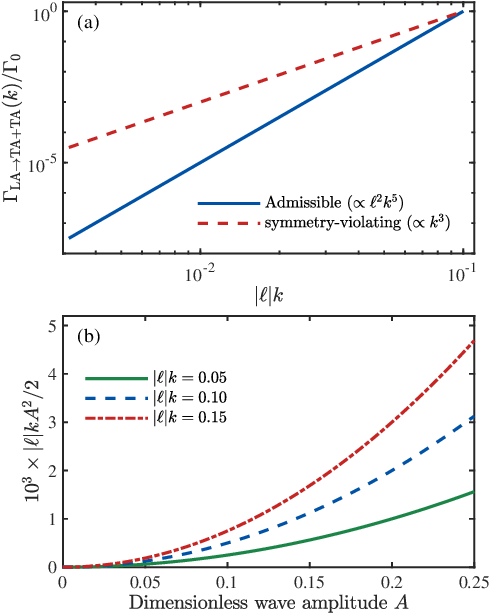}
\caption{(a) Normalized infrared decay rate $\Gamma_{\mathrm{LTT}}(k)/\Gamma_0$ [$\Gamma_0 \equiv \Gamma_{\mathrm{LTT}}(|\ell|k=0.1)$] for the symmetry-allowed vertex ($\propto \ell^2 k^5$, solid line) versus the illustrative shift-violating coupling ($\propto k^3$, dashed line). (b) Second-harmonic amplitude $|\ell|k A^2/2$ [\cref{eq:material_1d_stokes_shift}] of longitudinal Stokes waves governed by $\mathcal{L}_{\mathrm{1D}}^{\mathrm{adm}}$ [\cref{eq:material_1d_exact_lagrangian}] for representative values of $|\ell|k$.}
\label{fig:scaling_stokes}
\end{figure}

Kinematic accessibility alone does not establish that the decay occurs. The dynamical test is supplied by the nonvanishing $\mathrm{LTT}$ matrix element of the three-derivative cubic interaction in \cref{eq:material_complete_cubic}. As derived in Sec.~\ref{sec:cubic_interactions} and Appendix~\ref{app:decay_scaling}, combining this vertex with the three-dimensional final-state phase space yields the infrared rate $\Gamma_{\mathrm{LTT}} \propto \ell^2 k^5$. Figure~\ref{fig:scaling_stokes}(a) displays this scaling alongside the illustrative $k^3$ behavior of a shift-violating two-derivative coupling. While the $k^5$ scaling is characteristic of a derivative cubic phonon vertex and is not by itself unique to the present gauge-based construction, the model-specific content resides in the correlated dispersion relations and interaction structure.

We next consider a complementary nonlinear test: the propagation of longitudinal periodic waves. In one dimension, the closed-form action $\mathcal{L}_{\mathrm{1D}}^{\mathrm{adm}}$ [\cref{eq:material_1d_exact_lagrangian}] depends solely on derivatives of $R_{\hat{1}}{}^0$ and is exactly invariant under $R_{\hat{1}}{}^0 \to R_{\hat{1}}{}^0 + c$ at every nonlinear order. Consequently, its Euler--Lagrange equation takes conservation form, with no source term in its period average; by contrast, the unprojected constrained action can generate nonderivative contributions to that average. Solving the resulting nonlinear equation via the Stokes expansion---rather than invoking shift symmetry alone---yields the second harmonic in \cref{eq:material_1d_stokes_shift} with no frequency correction at order $A^2$. Figure~\ref{fig:scaling_stokes}(b) shows the corresponding second-harmonic amplitude $|\ell|k A^2/2$. This independent test probes the resummed longitudinal nonlinearities of the selected acoustic action.

The physical interpretation of these predictions rests on the specified material projection. The construction describes acoustic excitations about a fixed reference configuration of an elastic crystal, rather than gravitational interactions; the vierbein and induced metric encode an effective material geometry. The unrestricted commutator action [\cref{eq:commutator_action}] supplies the field-strength structure, whereas the kinematic constraints [\cref{eq:material_R_constraints}] isolate three independent acoustic fields and relate the spatial gauge potentials to the linearized symmetric strain. Restricting the field-strength contraction to its temporal and symmetric spatial sectors [\cref{eq:material_projected_action}] represents a minimal constitutive choice that eliminates quadratic restoring energy under infinitesimal rigid rotations. These restrictions reflect the physical response of the crystalline medium rather than constraints on a general gravitational gauge field.

In conventional Goldstone effective theories, uniform shifts constrain interactions modulo total derivatives~\cite{1994Leutwyler_PRD,1997Leutwyler_HEPA,2010Brauner_SYM}. The additional symmetry-admissibility prescription of Sec.~\ref{sec:cubic_interactions} implements a more restrictive selection by eliminating undifferentiated acoustic-field couplings---including those entering via the coframe and metric factors---while enforcing spacetime-inversion parity. It specifies a particular material interaction sector rather than the unique effective action allowed by the symmetries of an isotropic crystal. Once this prescription is fixed, the retained harmonic, cubic, and quartic terms, together with their relative coefficients, follow directly from the parent commutator action and its geometric factors without introducing vertex-dependent coupling constants. Their common dependence on $\ell$ thus establishes model-specific relations among dispersion, decay, and nonlinear wave response. Finite rotational objectivity and anisotropic material responses lie beyond the present linearized constitutive formulation.

Finally, the local gauge covariance of the parent construction must be distinguished from the physical translation symmetry realized in the acoustic sector. In the unrestricted theory, $D_{\hat{\alpha}} \to U D_{\hat{\alpha}} U^{-1}$ entails $G_{\hat{\alpha}\hat{\beta}} \to U G_{\hat{\alpha}\hat{\beta}} U^{-1}$, leaving the action invariant provided the metric and integration measure transform consistently. The material constraints, however, are imposed with respect to a fixed crystalline reference configuration. Although general local transformations acting on the gauge potentials while holding that reference fixed need not preserve the explicit form of these constraints, this feature does not spoil the gauge covariance of the parent action. The crystalline ground state spontaneously breaks continuous spatial translations, and the corresponding acoustic Goldstone modes are governed by an action invariant under uniform shifts $R_{\hat{i}}{}^0 \to R_{\hat{i}}{}^0 + c_{\hat{i}}$. These shifts represent uniform translations of the lattice and must be distinguished from both the local gauge transformations of the parent construction and arbitrary position-dependent shifts $c_{\hat{i}}(x)$. A manifestly gauge-covariant formulation of the material constraints, in which the crystalline reference structure transforms jointly with the gauge potentials, lies beyond the present fixed-reference description.


\section{Conclusions}
\label{sec:conclusions}

We have developed a symmetry-admissible acoustic sector of a nonlinear translational gauge construction for crystalline solids. Starting from the vierbein commutator action, the material kinematic constraints isolate three independent acoustic fields $R_{\hat{i}}{}^0$ and relate the spatial gauge potentials to the linearized symmetric strain. A minimal admissible contraction eliminates quadratic restoring energy under infinitesimal rigid rotations, while the subsequent shift- and inversion-symmetry prescription selects the derivative-based interactions admissible within the specified acoustic sector. Once these material conditions are fixed, the harmonic and retained nonlinear terms---together with their relative coefficients---follow directly from the parent commutator action and its geometric factors without introducing independent vertex couplings.

At harmonic order, the theory yields one longitudinal and two degenerate transverse gapless acoustic branches, whose predicted dispersions provide the kinematic basis for the subsequent analysis of phonon decay. At cubic order, the interaction yields a nonvanishing $\mathrm{LA}\to\mathrm{TA}+\mathrm{TA}$ matrix element and the infrared decay rate $\Gamma_{\mathrm{LTT}} \propto \ell^2 k^5$; the predicted dispersion branches keep this decay channel kinematically open across the entire longitudinal domain $|\ell|k < 1/2$. Independently, the closed-form longitudinal one-dimensional action supports periodic Stokes waves featuring a second harmonic and no leading frequency shift at order $A^2$.

This construction represents a material projection of a locally gauge-covariant parent theory, rather than a claim that the resulting acoustic action is invariant under unrestricted local translations. Relative to the fixed crystalline reference configuration, the acoustic action retains global spacetime-translation and uniform Goldstone-shift symmetries. Its predictions are therefore specific to the stated material and admissibility prescriptions, rather than universal consequences of translational gauge symmetry alone. Embedding the material reference structure into a fully gauge-covariant formulation, together with incorporating finite rotational objectivity and anisotropic constitutive responses, remains an open direction for future investigation.


\appendix

\section{Space-time components of the constrained field strength}
\label{app:field_strength_components}

Substituting the kinematic constraints of \cref{eq:material_R_constraints} into the general coordinate components of the translational field-strength tensor [\cref{eq:Gexpanded}], with $\partial_{\hat\alpha} = \delta_{\hat\alpha}{}^\mu\partial_\mu$, separates $\mathcal{G}_{\hat\alpha\hat\beta}{}^\mu$ into four space-time families entering the unrestricted gauge Lagrangian in \cref{eq:commutator_action}:
\begin{widetext}
\begin{equation}
\begin{aligned}
\mathcal{G}_{\hat0\hat i}{}^0
={}&-\mathrm{i}\hbar\partial_{\hat0}R_{\hat i}{}^0
 +\hbar^2
 R_{\hat j}{}^0\partial_{\hat j}R_{\hat i}{}^0,\\
\mathcal{G}_{\hat0\hat i}{}^j
={}&-\mathrm{i}\hbar(\partial_{\hat0}R_{\hat i}{}^j
 +\partial_{\hat i}R_{\hat j}{}^0)+\hbar^2\bigl(
 R_{\hat l}{}^0\partial_{\hat l}R_{\hat i}{}^j
 -R_{\hat i}{}^0\partial_{\hat0}R_{\hat j}{}^0
 -R_{\hat i}{}^l\partial_{\hat l}R_{\hat j}{}^0\bigr),\\
\mathcal{G}_{\hat i\hat j}{}^0
={}&-\mathrm{i}\hbar(\partial_{\hat i}R_{\hat j}{}^0
 -\partial_{\hat j}R_{\hat i}{}^0)-\hbar^2\bigl(
 R_{\hat i}{}^0\partial_{\hat0}R_{\hat j}{}^0
 -R_{\hat j}{}^0\partial_{\hat0}R_{\hat i}{}^0 +R_{\hat i}{}^l\partial_{\hat l}R_{\hat j}{}^0
 -R_{\hat j}{}^l\partial_{\hat l}R_{\hat i}{}^0\bigr),\\
\mathcal{G}_{\hat i\hat j}{}^k
={}&-\mathrm{i}\hbar(\partial_{\hat i}R_{\hat j}{}^k
 -\partial_{\hat j}R_{\hat i}{}^k)-\hbar^2\bigl(
 R_{\hat i}{}^0\partial_{\hat0}R_{\hat j}{}^k
 -R_{\hat j}{}^0\partial_{\hat0}R_{\hat i}{}^k +R_{\hat i}{}^l\partial_{\hat l}R_{\hat j}{}^k
 -R_{\hat j}{}^l\partial_{\hat l}R_{\hat i}{}^k\bigr).
\end{aligned}
\label{eq:material_G_components}
\end{equation}
\end{widetext}
These expressions explicitly display the separation used in Secs.~\ref{sec:effective_lagrangian} and~\ref{sec:cubic_interactions}. The component $\mathcal{G}_{\hat i\hat j}{}^0$ contains antisymmetric vorticity gradients $\partial_{\hat i}R_{\hat j}{}^0 - \partial_{\hat j}R_{\hat i}{}^0$ that give restoring energy to rigid rotations, while $\mathcal{G}_{\hat i\hat j}{}^k$ vanishes for a static infinitesimal rigid rotation and is excluded in the minimal truncation of $\mathcal{L}_{\mathrm{ph}}$ [\cref{eq:material_projected_action}]. Among the quadratic commutator terms ($\propto \hbar^2$) in $\mathcal{G}_{\hat0\hat i}{}^0$ and $\mathcal{G}_{\hat0\hat i}{}^j$, only $-\hbar^2 R_{\hat i}{}^l\partial_{\hat l}R_{\hat j}{}^0$ contains no undifferentiated $R_{\hat i}{}^0$ factors and survives the shift-invariance projection in \cref{eq:G_self_admissible}.


\section{Infrared $\mathrm{LA} \to \mathrm{TA} + \mathrm{TA}$ decay rate}
\label{app:decay_scaling}

To evaluate the zero-temperature decay rate in \cref{eq:decay_rate_golden_rule} explicitly, we first construct the cubic interaction Hamiltonian $\hat{H}_{\mathrm{ph}}^{(3)}$ from the complete three-derivative Lagrangian [\cref{eq:material_complete_cubic}],
\begin{equation}
\begin{aligned}
\mathcal{L}_{\mathrm{ph},D=3}^{(3)}
={}&\mathrm{i}\hbar^3\ell
(\partial_{\hat{l}}R_{\hat{l}}{}^0)
\Biggl[\sum_{\hat{i}}(\partial_{\hat{0}}R_{\hat{i}}{}^0)^2\\
&-\frac{1}{4}\sum_{\hat{i},\hat{j}}
(\partial_{\hat{i}}R_{\hat{j}}{}^0
+\partial_{\hat{j}}R_{\hat{i}}{}^0)^2\Biggr].
\end{aligned}
\label{eq:app_cubic_recalled}
\end{equation}
Because the harmonic Lagrangian $\mathcal{L}_{\mathrm{ph}}^{(2)}$ involves mixed space-time derivatives, the generalized canonical momentum conjugate to $R_{\hat{i}}{}^0$ (with $\partial_{\hat{0}}=c_s^{-1}\partial_t$) reads
\begin{equation}
\begin{aligned}
\Pi_{\hat{i}}={}&
\frac{\partial\mathcal{L}_{\mathrm{ph}}}
{\partial(\partial_{\hat{0}}R_{\hat{i}}{}^0)}
-\partial_{\hat{j}}\frac{\partial\mathcal{L}_{\mathrm{ph}}}
{\partial(\partial_{\hat{j}}\partial_{\hat{0}}R_{\hat{i}}{}^0)}\\
={}&-\hbar^2\partial_{\hat{0}}R_{\hat{i}}{}^0
+\hbar^2\ell\partial_{\hat{j}}
(1-2\ell\partial_{\hat{0}})
(\partial_{\hat{i}}R_{\hat{j}}{}^0
+\partial_{\hat{j}}R_{\hat{i}}{}^0)\\
&+2\mathrm{i}\hbar^3\ell
(\partial_{\hat{j}}R_{\hat{j}}{}^0)
\partial_{\hat{0}}R_{\hat{i}}{}^0+\cdots ,
\end{aligned}
\label{eq:app_generalized_momentum}
\end{equation}
where the ellipsis denotes higher-derivative terms. Performing the Legendre transformation $H_{\mathrm{ph}}=c_s\int \mathrm{d}^3x\,[\Pi_{\hat{i}}\partial_{\hat{0}}R_{\hat{i}}{}^0-\mathcal{L}_{\mathrm{ph}}]$ to cubic order in the interaction picture yields
\begin{equation}
\hat{H}_{\mathrm{ph}}^{(3)}
=-c_s\int \mathrm{d}^3x\,
\mathcal{L}_{\mathrm{ph},D=3}^{(3)}\big|_{\mathrm{free}},
\label{eq:app_cubic_Hamiltonian}
\end{equation}
where the time derivatives are evaluated on the free fields. Quantizing the harmonic sector in a volume $V$, we expand the field operator as
\begin{equation}
\begin{aligned}
\hat{R}_{\hat{i}}{}^0={}&\frac{\mathrm{i}\mathcal{N}}{\sqrt{V}}
\sum_{\bm{q},s}\frac{\hat{e}_{\hat{i}}^{(s)}(\bm{q})}
{\sqrt{\omega_s(q)Z_s(q)}}\\
&\times\Bigl[\hat{a}_{\bm{q},s}
 e^{\mathrm{i}(\bm{q}\cdot\bm{x}-\omega_s t)}
+\hat{a}_{\bm{q},s}^{\dagger}
 e^{-\mathrm{i}(\bm{q}\cdot\bm{x}-\omega_s t)}\Bigr].
\end{aligned}
\label{eq:app_R_modes}
\end{equation}
Here $\hat{\bm{e}}^{(s)}(\bm{q})$ are unit polarization vectors, while $Z_{\mathrm{LA}}(q)=1-4\ell^2q^2$ and $Z_{\mathrm{TA}}(q)=1-2\ell^2q^2$ denote the kinetic renormalization factors. The mode normalization constant $\mathcal{N}$ is fixed by canonical quantization once the overall dimensional prefactor of the gauge action is specified; because this prefactor is left unfixed in the present effective formulation, the absolute prefactor of the decay rate remains undetermined, whereas its infrared $k^5$ scaling is robust.

For the decay process $\ket{i}=\ket{\mathrm{LA}(\bm{k})} \to \ket{f}=\ket{\mathrm{TA}(\bm{q}_+,\sigma_+),\mathrm{TA}(\bm{q}_-,\sigma_-)}$ with $\bm{q}_- = \bm{k} - \bm{q}_+$, only the incoming longitudinal mode contributes to the divergence $\partial_{\hat{l}}R_{\hat{l}}{}^0$ in \cref{eq:app_cubic_recalled}. Contracting the remaining two fields with the outgoing transverse modes (including both final-state permutations) yields the unnormalized three-phonon vertex amplitude
\begin{equation}
\begin{aligned}
\mathcal{V}_{\sigma_+\sigma_-}^{(3)}
={}&-\hbar^3\ell k\,\mathcal{B}_{\sigma_+\sigma_-},\\
\mathcal{B}_{\sigma_+\sigma_-}={}&
\biggl[\bm{q}_+\cdot\bm{q}_-
-\frac{2\omega_{\mathrm{TA}}(q_+)
\omega_{\mathrm{TA}}(q_-)}{c_s^2}\biggr]
(\hat{\bm{e}}_+\cdot\hat{\bm{e}}_-)\\
&+(\bm{q}_+\cdot\hat{\bm{e}}_-)
(\bm{q}_-\cdot\hat{\bm{e}}_+),
\end{aligned}
\label{eq:app_LTT_amplitude}
\end{equation}
which receives contributions from both the temporal and spatial sectors of \cref{eq:app_cubic_recalled} and reduces to $\mathcal{V}_{\sigma_+\sigma_-}^{(3)} = \hbar^3\ell k^3/2$ in the orthogonal decay geometry considered in Sec.~\ref{sec:cubic_interactions}.

To extract the long-wavelength scaling, we introduce the dimensionless wave vectors $\bm{p}_+ = \bm{p} = \bm{q}_+/k$ and $\bm{p}_- = \hat{\bm{k}} - \bm{p} = \bm{q}_-/k$. Expanding \cref{eq:material_linear_dispersions} for $|\ell|k \ll 1$ gives $\omega_{\mathrm{LA}}(k) = c_s k + \mathcal{O}(c_s\ell^2 k^3)$ and $\omega_{\mathrm{TA}}(q_\pm) = c_s k p_\pm/\sqrt{2} + \mathcal{O}(c_s\ell^2 k^3)$, together with $Z_s(q) = 1 + \mathcal{O}(\ell^2 k^2)$. Consequently, $\mathcal{B}_{\sigma_+\sigma_-} = k^2 b_{\sigma_+\sigma_-} + \mathcal{O}(\ell^2 k^4)$, with the dimensionless kinematic and polarization factor
\begin{equation}
\begin{aligned}
b_{\sigma_+\sigma_-}={}&
(\bm{p}_+\cdot\bm{p}_--p_+p_-)
(\hat{\bm{e}}_+\cdot\hat{\bm{e}}_-)\\
&+(\bm{p}_+\cdot\hat{\bm{e}}_-)
(\bm{p}_-\cdot\hat{\bm{e}}_+).
\end{aligned}
\label{eq:app_b_dimensionless}
\end{equation}
Substituting \cref{eq:app_R_modes,eq:app_LTT_amplitude,eq:app_b_dimensionless} into \cref{eq:app_cubic_Hamiltonian}, the squared transition matrix element evaluates to
\begin{equation}
\begin{aligned}
\bigl|\bra{f}\hat{H}_{\mathrm{ph}}^{(3)}\ket{i}\bigr|^2
={}&\frac{2\mathcal{N}^6\hbar^6}{c_sV}
\ell^2k^3\,
\frac{|b_{\sigma_+\sigma_-}|^2}{p_+p_-}\\
&\times\bigl[1+\mathcal{O}(\ell^2 k^2)\bigr].
\end{aligned}
\label{eq:app_M_squared}
\end{equation}

Finally, in the continuum limit $\sum_{\bm{q}_+} \to [V k^3/(2\pi)^3] \int \mathrm{d}^3p$, the energy-conserving delta function in \cref{eq:decay_rate_golden_rule} expands as
\begin{equation}
\delta(E_i-E_f)=\frac{1}{\hbar c_s k}
\delta\!\left(1-\frac{p_++p_-}{\sqrt{2}}\right)
+o(k^{-1}).
\label{eq:app_phase_space}
\end{equation}
Introducing spherical coordinates $(p,\mu,\phi)$ aligned with $\hat{\bm{k}}$, where $p_+ = p$, $\mu = \hat{\bm{k}}\cdot\hat{\bm{p}}$, and $p_- = \sqrt{1+p^2-2p\mu}$, the delta function enforces $\mu = \sqrt{2} - 1/(2p)$ within the kinematic interval $(\sqrt{2}-1)/2 \le p \le (\sqrt{2}+1)/2$. Its Jacobian $\sqrt{2}\,p_-/p$ combines with the spherical volume element $p^2\,\mathrm{d}p\,\mathrm{d}\mu\,\mathrm{d}\phi$ to cancel the $(p_+ p_-)^{-1}$ factor in \cref{eq:app_M_squared}. After summing over the transverse polarizations, rotational invariance about $\hat{\bm{k}}$ renders the integrand independent of $\phi$, so that the azimuthal integration yields the $k$-independent, strictly positive dimensionless constant
\begin{equation}
\mathcal{I} = 2\sqrt{2}\pi\int_{(\sqrt{2}-1)/2}^{(\sqrt{2}+1)/2}
\mathrm{d}p\sum_{\sigma_+,\sigma_-}|b_{\sigma_+\sigma_-}|^2 > 0
\label{eq:app_angular_integral}
\end{equation}
(for instance, out-of-plane polarizations at $\theta(0)=45^\circ$ yield $b = -1/2$). Combining \cref{eq:app_M_squared,eq:app_phase_space,eq:app_angular_integral} in \cref{eq:decay_rate_golden_rule} gives the infrared decay rate
\begin{equation}
\begin{aligned}
\Gamma_{\mathrm{LTT}}(k)
={}&\frac{\pi}{\hbar}\,
\frac{2\mathcal{N}^6\hbar^6\ell^2k^3}{c_sV}
\frac{Vk^3}{(2\pi)^3}\,
\frac{\mathcal{I}}{\hbar c_s k} + o(k^5)\\
={}&\frac{\mathcal{N}^6\hbar^4}{4\pi^2c_s^2}
\mathcal{I}\,\ell^2k^5+o(k^5).
\end{aligned}
\label{eq:app_golden_rescaled}
\end{equation}

\clearpage



%

\end{document}